\documentclass[epj]{svjour}
\usepackage{graphicx}% Include figure files
\begin{document}
\title{
%\hfill{\small {\bf MKPH-T-03-03}}\\
\bf Final state interaction in spin asymmetry and GDH
sum rule for incoherent pion production on the deuteron
} 
\author{E.M.\ Darwish\thanks{Present address: Physics Department,
Faculty of Science, South Valley University, Sohag, Egypt.}, 
H.\ Arenh\"ovel, and M.\ Schwamb}
\institute{
Institut f\"{u}r Kernphysik, Johannes Gutenberg-Universit\"{a}t,
       D-55099 Mainz, Germany}
%\date{Received: date / Revised version: date}
\date{\today, preprint MKPH-T-03-03 }
\abstract{ 
The contribution of incoherent single pion photoproduction to the spin
response of the deuteron, i.e., the asymmetry of the total
photoabsorption cross section 
with respect to parallel and antiparallel spins of photon and deuteron, is 
calculated over the region of the $\Delta$-resonance with inclusion of
final state $NN$- and $\pi N$-rescattering. Sizeable effects, mainly from 
$NN$-rescattering, are found leading to an appreciable reduction
of the spin asymmetry. Furthermore, the contribution to 
the Gerasimov-Drell-Hearn integral is explicitly evaluated by integration 
up to a photon energy of 550 MeV. Final state interaction reduces the 
value of the integral to about half of the value obtained for the pure
impulse approximation. 
}

\titlerunning{FSI in spin asymmetry of pion production on deuteron}
\maketitle

\section{Introduction}
\label{sec1}

The spin response or vector target asymmetry of the total photo
absorption cross section of a hadron or nucleus has come into 
focus~\cite{Dre95,Proc01} with recent interest in the 
Gerasimov-Drell-Hearn (GDH) sum rule~\cite{Ger65,DrH66}. This 
sum rule connects the anomalous magnetic moment  $\kappa$ of a
particle of mass $M$, charge $eQ$, and spin $S$ with the energy
weighted integral over the difference 
$\sigma ^P(k)-\sigma^A(k)$ of the total photoabsorption cross 
sections for circularly polarized photons on a target with spin
parallel and antiparallel to the spin of the photon
\begin{equation}
I^{GDH}(\infty)=\int_0^\infty \frac{dk}{k}
\left(\sigma ^P(k)-\sigma ^A(k)\right)
=4\pi^2\kappa^2\frac{e^2}{M^2}\,S
\,.\label{gdh}
\end{equation}
Here the anomalous magnetic moment is 
defined by the total magnetic moment operator of the particle 
$\vec M = e\,(Q+\kappa)\,\vec S/M$, where $\vec S$ denotes the
spin operator of the target.

For the deuteron, one finds a very small anomalous magnetic moment
$\kappa_d=-0.143$ resulting in a GDH prediction of $I^{GDH}_d =
0.65\,\mu$b, which is more than two orders of magnitude smaller 
than the nucleon values, i.e., $I^{GDH}_p=204.8\,\mu$b 
for the proton and $I^{GDH}_n=233.2\,\mu$b for the neutron. The
explicit evaluation of the spin asymmetry for various reaction
channels, essentially photodisintegration and coherent and incoherent
single pion production, has shown, that this small GDH sum rule value
is the result of large cancellations between the contributions of the 
individual channels~\cite{ArK97}. For photodisintegration alone one finds
a large negative spin asymmetry close to the break-up threshold which
arises from the large isovector $M1$-transition to the resonant 
$^1S_0$-state. In fact, the explicit evaluation of the GDH integral
for photodisintegration 
up to an energy of 550 MeV yields a large negative contribution of
$I^{GDH}_{\gamma d \to np}(550\,\mbox{MeV})=-413\,\mu$b, which almost
equals the sum of the GDH values of proton and neutron. We would like
to remark that for this channel relativistic contributions were 
quite sizeable already at rather low energies below 50 MeV, 
reducing the GDH integral from $-665~\mu$b by about one third 
to the foregoing value. Furthermore, the GDH integral had already 
reached convergence at 550 MeV. In view of this negative contribution
a corresponding positive contribution has to be expected from the other
channels, i.e., mainly from pion photoproduction. 

In this context it has been suggested to `measure' in the absence of
neutron targets the neutron spin asymmetry by measuring the spin
asymmetry of the deuteron for the pion photoproduction channels using
a vector polarized deuteron target, thus allowing the explicit
evaluation of the GDH integral for the neutron. The implicit tacit 
assumption underlying this suggestion is that the contributions of 
pion production on proton and neutron in the deuteron add up incoherently. 
As was shown in~\cite{ArK97}, already for the impulse approximation 
this is not the case. With the present note we will demonstrate that
it is even less fulfilled if final state interaction effects are included. 

Notwithstanding the
`caveats' of such an `extraction`~\cite{Are00}, it is certainly of
interest to study the deuteron spin asymmetry for the pion production
channels, because in general polarization observables will give additional 
valuable information for checking the spin degrees of freedom of the 
elementary pion production amplitude of the neutron, provided, and
this is very important, that one has under control all interfering
interaction effects which prevent a simple extraction of this amplitude. 

\section{Final state interaction in incoherent pion production on 
the deuteron}
\label{sec3c}
In our previous evaluation~\cite{ArK97}, the incoherent pion production
contribution to spin asymmetry and GDH integral had been evaluated in
the pure impulse approximation (IA) only, i.e., neglecting any final state
interaction effects (FSI) and possible two-body contributions to the
production operator. In this framework, the reaction proceeds via the
pion production on one nucleon while the other nucleon acts merely as
a spectator. In view of the dominance of
the quasifree production process one could expect that the IA would
work reasonably well for charged pion production. However, for neutral
pion production there is some double counting with respect to the
coherent process due to the neglect of $NN$-rescattering in the final
state with the effect that the final state is not orthogonal to the
$^3S_1$-$^3D_1$ state  of the deuteron. Thus in this case one expects
a sizeable overestimation by the IA, because it contains a fraction 
of the coherent channel. This has been confirmed
in~\cite{LeS00,DaA03} where FSI effects were considered, and a sizeable
reduction of the unpolarized 
total and differential cross sections of the incoherent
neutral pion production had been found due to the above mentioned 
non-orthogonality, whereas for charged pion production the FSI effects 
were significantly smaller.

\begin{figure}[htb]
\resizebox{0.49\textwidth}{!}{%
\includegraphics{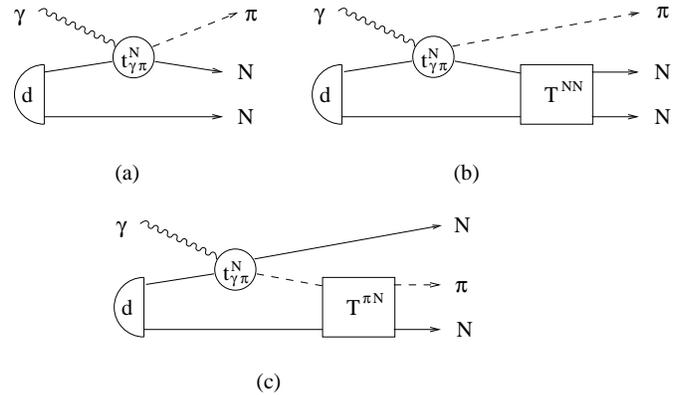}
}
\caption{Diagrammatic representation of $\gamma d\rightarrow \pi NN$ 
including rescattering in the two-body subsystems of the final state: 
(a) impulse approximation (IA), (b) $NN$-rescattering, and (c) 
$\pi N$-rescattering.} 
\label{t-matrix}
\end{figure}

In~\cite{DaA03} we have included besides the pure
impulse approximation the complete rescattering by the final
state interaction within each of the two-body subsystems $NN$- and $\pi
N$. For the pion
the rescattering refers to the scattering on the spectator nucleon 
because the rescattering on the active nucleon is already included in
the dominant elementary $M_{1+}^{3/2}$ multipole. Therefore, the total
transition matrix element reads in this approximation 
\begin{eqnarray}
\label{three}
{\cal M}^{(t\mu)} & = &{\cal M}^{(t\mu)~IA} + 
{\cal M}^{(t\mu)~NN} + {\cal M}^{(t\mu)~\pi N}\,  
\end{eqnarray}
in an obvious notation. A graphical representation of the scattering
matrix is shown in Fig.~\ref{t-matrix}. The necessary half-off-shell
two-body scattering matrices were obtained from separable
representations of realistic $NN$- and $\pi N$-interactions which gave
good descriptions of the corresponding phase shifts~\cite{Dar02}. For
$NN$-rescattering, we have included all partial waves with total
angular momentum $J\le 3$ and for $\pi N$-rescattering $S$- through
$D$-waves. As $NN$-interaction we have used the Paris potential, but
we have also tried the Bonn r-space potential obtaining essentially
the same result. For further details with respect to the rescattering
contributions we refer to~\cite{DaA03}.
For the elementary pion photoproduction operator, we have taken as 
in our previous work the standard pseudovector Born terms and the 
contribution of the $\Delta$-resonance as described in detail 
in~\cite{ScA96}, and a satisfactory
description of pion photoproduction on the nucleon in the
$\Delta$-resonance region~\cite{DaA03} was achieved. 

The major effect of $NN$- and $\pi N$-rescattering in incoherent pion
photoproduction on the deuteron was a reduction of the total unpolarized 
cross section which amounts 
in the maximum of the $\Delta$(1232)-resonance region for charged pion
photoproduction to about 5 percent and to about 35 percent for 
$\pi^0$-photoproduction. Thus for charged pion production the IA 
appears to be a reasonable approximation within these five percent, 
but not for neutral pion production as already
mentioned above. The dominant effect of FSI came from $NN$-rescattering
whereas $\pi N$-rescattering was much less important, almost
negligible (compare the dashed curves with the solid ones in 
Fig.~11 of~\cite{DaA03} which are almost indistinguishable). With respect to 
experimental data a satisfactory agreement was achieved. 

\section{Results for spin asymmetry and GDH integral}

\begin{figure}[htb]
\resizebox{0.49\textwidth}{!}{%
\includegraphics{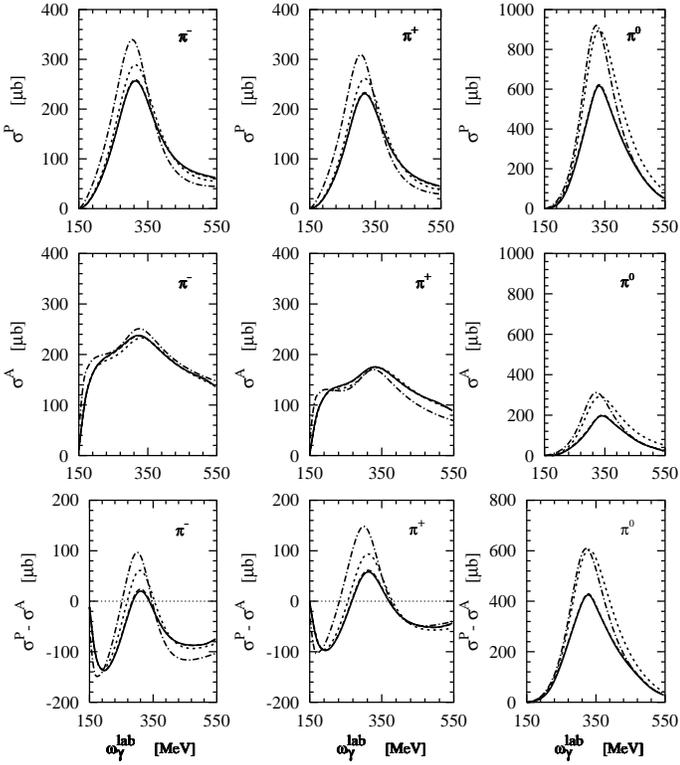}}
\vspace*{.5cm}
\caption{Total absorption cross sections for circularly polarized
photons on a target with spin parallel $\sigma ^{P}$ (upper part) and
antiparallel $\sigma ^{A}$ (middle part) to the photon spin. Lower part
shows the difference $\sigma ^{P}-\sigma ^{A}$. Notation: dotted
curves: impulse approximation (IA); dashed curves: IA+$NN$
rescattering; solid curves: IA+$NN$+$\pi N$
rescattering; dash-dotted curves: cross sections for
$\pi^-$ production on the neutron (left panels), $\pi^+$ on the proton
(middle panels) and $\pi^0$ on both, proton and neutron (right panels).}
\label{spin_asym_all}
\end{figure}

In the meantime, we have evaluated also the spin asymmetry for these 
reactions. The results are presented in Fig.~\ref{spin_asym_all}, 
where the upper
part shows the total photoabsorption cross sections $\sigma ^{P}$ for
circularly polarized photons on a target with spin parallel to the
photon spin, the middle part $\sigma ^{A}$, the one for antiparallel
spins of photon and target and the lower part the spin asymmetry
$\sigma^{P} - \sigma ^{A}$ for the individual contributions of the
different pion 
charge states to incoherent pion photoproduction on the deuteron. 
For comparison, we also show in the same figure the results on the
free nucleon by the dash-dotted curves. In the case of $\pi^0$ it is
the sum of the proton and neutron asymmetries. 

One notes for the cross sections $\sigma^{P}$ and 
$\sigma^{A}$ as well as for the spin asymmetry of the nucleons 
and the deuteron qualitatively a similar behaviour, although for
the deuteron the maxima and minima are smaller and also slightly
shifted towards higher energies. Furthermore, in $\sigma^{P}$ the
charged pion contributions show a larger deviation between the IA and
the elementary one, whereas for $\sigma^{A}$ the differences are smaller. 
In contrast to this, one notes for the neutral pion contributions a
much closer behaviour between the IA and the elementary reaction for 
both cross sections and the difference. 

FSI effects appear for charged pion production 
mainly in $\sigma^{P}$ while for neutral pions they are of equal importance 
for both $\sigma^{P}$ and $\sigma^{A}$, again because of the non-orthogonal 
final state in IA. The bottom panels in Fig.~\ref{spin_asym_all} show that 
final state interactions lead to an overall strong reduction of the spin 
asymmetry in the energy region of the $\Delta$(1232)-resonance. This reduction
becomes about 170~$\mu$b for $\pi^0$-production and about 35~$\mu$b for 
charged pions on the $\Delta$-peak. 
Thus even for charged pion production the IA is not 
anymore a reasonable approximation as it was for the unpolarized total 
cross section. Moreover, already the IA deviates significantly from the 
corresponding nucleon quantities for charged pion production. These facts
underline our doubts concerning an 'extraction' of the
neutron spin asymmetry from the one of the deuteron. 
It is also obvious that $\sigma^P$ is much
larger than $\sigma^A$ because of the $\Delta$-excitation, 
in particular in the case of $\pi^0$-production. The influence of FSI stems 
predominantly from $NN$-rescattering whereas $\pi N$-rescattering is much 
smaller. In fact, the dashed curves, representing the inclusion of $NN$-FSI 
alone, and the solid ones with both $NN$- and $\pi N$-FSI appear almost 
indistinguishable. It is interesting to note that the size of the reduction 
of $\sigma^P-\sigma^A$ in the IA by FSI for $\pi^0$-production 
(about 170~$\mu$b in the maximum), related to the already mentioned 
non-orthogonality effect, is comparable in magnitude, though somewhat 
larger, to the coherent contribution to the spin asymmetry of 120~$\mu$b 
in the maximum as reported in~\cite{ArK97}. 
\begin{figure}
\resizebox{0.49\textwidth}{!}{%
\includegraphics{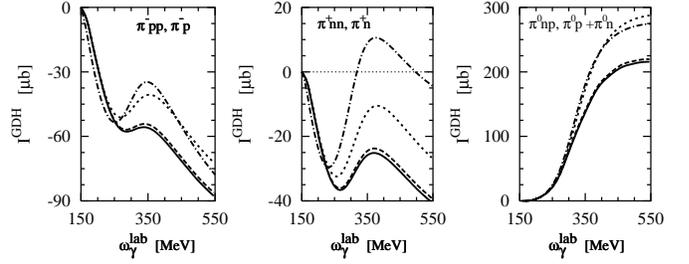}}
\vspace*{.5cm}
\caption{The Gerasimov-Drell-Hearn integral as a function of the upper
integration limit for the different channels of incoherent single
pion photoproduction on the deuteron and the nucleon. Notation of the
curves as in Fig.~\ref{spin_asym_all}.} 
\label{gdh_integral}
\end{figure}

\begin{figure}
\resizebox{0.45\textwidth}{!}{%
\includegraphics{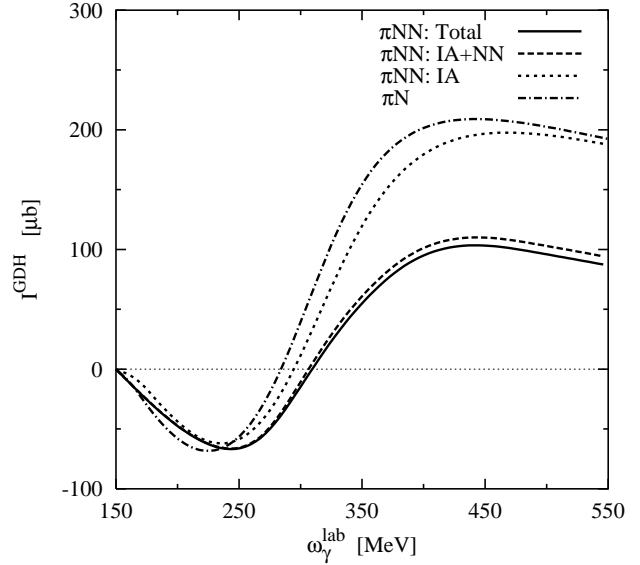}}
\vspace*{.5cm}
\caption{Total contribution of the three channels of
incoherent single pion photoproduction to the GDH integral for the
deuteron and the nucleon as function of the upper integration
limit. Notation of the curves as in Fig.~\ref{spin_asym_all}.}
\label{gdh_integral_sum}
\end{figure}

In Fig.~\ref{gdh_integral} the corresponding GDH integrals are shown for 
the separate channels and in Fig.~\ref{gdh_integral_sum} for the sum of 
all three pion channels as function of the upper integration limit. One 
notes again the significant reduction by FSI, mainly through 
$NN$-rescattering, but the $\pi N$-rescattering effect is now
distinguishable although still quite small 
(compare the dashed with the solid curves in 
Figs.~\ref{gdh_integral} and \ref{gdh_integral_sum}). 
\begin{table}[h]
\caption{\small Contributions of incoherent single pion photoproduction to
  the GDH integral for the deuteron integrated up to 550 MeV in $\mu$b.}
\begin{center}
\renewcommand{\arraystretch}{1.5}
\begin{tabular}{|c|ccc|}
\hline
reaction&$I^{GDH}_{IA}$&$I^{GDH}_{IA+NN}$&$I^{GDH}_{IA+NN+\pi N}$\\ 
\hline
$\gamma d \to pp\pi^-$ &   $-73$   & $-87$   & $-88$ \\    
$\gamma d \to nn\pi^+$ &   $-27$   & $-39$   & $-41$ \\
$\gamma d \to np\pi^0$ &   287   & 220   & 216 \\ 
\hline
$\gamma d \to \pi NN$                  &   187   & 94    & 87  \\
\hline
 \end{tabular}
\end{center}
\label{tab1}
\end{table}

The values of the GDH integral up 
to 550 MeV for all three channels and their sum are listed in 
Table~\ref{tab1}. A sizeable positive 
contribution comes from $\pi^0$-production whereas the charged pions give a 
negative but - in absolute size - smaller contribution to the GDH value. 
The inclusion of $NN$-FSI reduces the total GDH integral to one half of 
the IA value which is then further reduced, but on a much smaller scale 
by $\pi N$-rescattering. Furthermore, we would like to point out that the 
total value of the integral of 87 $\mu$b is thus considerably smaller 
than the sum of the neutron and proton values for the present model of the 
elementary model, namely 
$I^{GDH}_n(550\,\mbox{MeV})+I^{GDH}_p(550\,\mbox{MeV})=197\,\mu$b. This
underlines our caveat, that a simple extraction of the neutron spin 
asymmetry from a measurement on the deuteron is not possible.

\section{Concluding remarks}
\label{sec4}
If one adds the previously reported values of the GDH
integral from the  
photodisintegration channel 
($I^{GDH}_{\gamma d\rightarrow np}(550\,\mbox{MeV})=-413\,\mu$b) 
and the coherent pion production 
($I^{GDH}_{\gamma d\rightarrow \pi^0 d}(550\,\mbox{MeV})=63\,\mu$b)
from~\cite{ArK97}, one finds for the total GDH value from explicit 
integration up to 550 MeV a negative value
$I^{GDH}_d(550\,\mbox{MeV})=-263\,\mu$b which has to be compared to the 
value $I^{GDH}_d(\infty)=0.65\,\mu$b. However, as was already
mentioned in~\cite{ArK97}, the model of the elementary production
amplitude above the $\Delta$-resonance is not very realistic, because
it had been constructed primarily to give a realistic description of
the $\Delta$-resonance region and not at higher energies. In fact,
it grossly underestimates the GDH integral up to 550 MeV if we compare
$I^{GDH}_{n+p}(550\,\mbox{MeV})=197\,\mu$b of our model with the most 
recent MAID 2000 analysis~\cite{Maid} yielding a value 
$I^{GDH}_{n+p}(550\,\mbox{MeV})=282\,\mu$b. The latter value is 
based on a multipole analysis of experimental pion photoproduction data. 
Thus it is clear that the large negative value for the deuteron 
will be reduced significantly 
using an elementary production operator which yields a better 
description above the $\Delta$-resonance. Moreover, the pion 
production contribution has not yet reached convergence at 550 MeV 
(see Figs.~\ref{gdh_integral} and \ref{gdh_integral_sum}).
In fact, one expects at energies above 550 MeV further positive
contributions from pion production~\cite{Tia02}. Thus for a more quantitative
evaluation one has to use a more realistic elementary operator. In
addition, an appropriate $NN$-interaction model which includes retardation
effects and inelasticities is needed as well as the consideration of
relativistic effects which turned out to be very important for the spin
asymmetry of the photodisintegration channel~\cite{ArK97} as mentioned above. 
In addition, a genuine three-body approach would be desirable.

Thus we would like to conclude that, notwithstanding the shortcomings
of the present work, the results clearly show 
first of all the importance of final state interaction effects in the 
spin asymmetry of the deuteron, a feature which certainly will remain true
in a more realistic treatment of the energy region above the 
$\Delta$-resonance. Secondly, it is obvious that 
a measurement on the deuteron will not provide direct 
access to the spin asymmetry of the neutron. However, the deuteron
spin asymmetry will certainly provide us with an important check observable
for testing our knowledge of the pion photoproduction on the neutron and
thus will give us valuable information on the neutron's spin asymmetry in 
an indirect manner if one has a reliable model for the FSI and possible other
two-body effects at hand for the analysis. 

\section*{Acknowledgement} 
This work was supported by the Deutsche Forschungsgemeinschaft (SFB 443).
E.M.\ Darwish acknowledges a fellowship from Deutscher 
Akademischer Austauschdienst (DAAD).

\end{document}